\documentclass[graybox,natbib,nosecnum]{svmult}
\bibpunct{(}{)}{;}{a}{}{,} 

\pdfoutput=1   

\usepackage{mathptmx}       
\usepackage{helvet}         
\usepackage{courier}        
\usepackage{type1cm}        

\usepackage{makeidx}         
\usepackage{graphicx}        
\usepackage{multicol}        
\usepackage[bottom]{footmisc}
\usepackage[normalem]{ulem}	
\usepackage{hyperref}  

\usepackage{soul}   

\makeindex             

\begin{document}

\title*{Building a Planet Atmosphere: Fundamental Physics and Chemistry}
\author{Emily Rauscher}
\institute{Emily Rauscher \at Department of Astronomy, University of Michigan, 1085 S. University Ave., Ann Arbor, MI 48109, USA, \email{erausche@umich.edu}}

%
%
\maketitle

\abstract{This chapter provides an overview of the basic concepts foundational to atmospheric physics and chemistry. We discuss the retention of atmospheres against thermal evaporation and the global energy balance of planets. We present simple derivations of the vertical profile of an atmosphere, which may be shaped by convective and radiative transport. We then briefly touch upon the three-dimensional atmospheric structure, as shaped by circulation patterns. We describe how the abundances of chemical species in the atmosphere are determined, starting with the assumption of chemical equilibrium and then expanding to various disequilibrium effects. We introduce the particles that can be important components of atmospheres (clouds and hazes) and sketch out some of their complexity. Finally, we review some of the differences between atmospheres of terrestrial and gaseous worlds. }

\section{Introduction}

The definition of a planet does not require that it maintain any substantial atmosphere, with Mercury and some exoplanets as examples of effectively bare rock worlds \citep[e.g., LHS 3844b and TRAPPIST-1 b;][]{Kreidberg2019,Greene2023}. However, for those planets that are able to maintain a gravitationally bound layer of gas, this atmosphere acts as the interface between the planet and the rest of the universe. The atmosphere filters the starlight incident on the planet and moderates the light that leaves the planet. Terrestrial worlds have atmospheres that sit atop a solid surface, while the atmosphere of a gas giant is simply the outermost region of its gaseous envelope (which encompasses most of the planet's mass). Regardless of the type of planet, its atmospheric properties will be shaped by its formation and evolutionary history and can therefore provide clues into those mechanisms. Atmospheres also host a range of physical and chemical processes; the diversity of conditions realized among exoplanets offers the opportunity to extend our understanding of these processes into new and extreme regimes. We are often particularly interested in those atmospheres closest to being like Earth, to look for evidence of habitable conditions and even hints of life. Atmospheres present a rich tapestry that connects clues of formation and habitability with fundamental physical and chemical processes.

This review is meant as a primer on the basics of atmospheric physics and chemistry. We touch on the various processes that govern the properties and behavior of atmospheres, with the goal of helping the reader to achieve a conceptual understanding of the fundamentals. Much of the material here is covered in more detail in standard textbooks. We highlight \cite{Pierrehumbert} and \cite{Seager} as good references for this material, which focus on Solar System and exoplanet atmospheres, respectively, but there are many other great texts available. When we touch on atmospheric applications of physical and chemical processes, or details that may not be as easily found in textbooks, we provide references to example cases. However, as the goal of this chapter is a basic understanding, we are not comprehensively reviewing the literature to highlight the latest work. We do encourage the reader to explore the cross-referenced chapters (listed at the end), which contain complementary information and more detailed discussions of some topics.

\section{Atmospheric Physics}

We begin with the physical foundations of an atmosphere being set by its two defining properties: 1) it is a layer of gas bound to the planet and 2) it is the medium through which starlight is absorbed or reflected and the planet's thermal emission emerges. This means that we can determine the properties and structure of a planet's atmosphere using the physics of: gas properties, gravity, global energy balance, and heat transport (through convection or radiation).

\subsection{The role of gravity: bound atmospheres and pressure profiles}

For gas to be bound to a planet, gravity needs to overcome any forces working to evaporate away the material. In this context, we are using ``evaporation" to describe a situation where the temperature of atmospheric gas is high enough for it to become gravitationally unbound from the planet. (Later in the chapter we will use ``evaporation" differently, to refer to the phase change of material from a liquid to gaseous state.) The most basic criterion for stability against evaporation is that the escape velocity of the planet must be many times greater than the thermal velocity of the gas. In thermodynamic equilibrium the particles in a gas will have a range of speeds set by the Maxwell-Boltzmann distribution and if some significant fraction of the high-velocity tail exceeds the escape speed of the planet, then those gas particles with upwards trajectories will fly away, unbound from the planet. In principle there is no upper bound to the Maxwell-Boltzmann distribution, but if a negligible enough fraction of gas particles reach velocities above the escape speed, then this effectively means that the evaporation timescale becomes infinitely long. While there is no formal limit as to when an atmosphere is safe against thermal evaporation, we often use a factor of 6 difference between the escape and thermal velocities for this criterion and say that the atmosphere is stable when $v_{\mathrm{esc}}>6v_{\mathrm{th}}$:
\begin{equation}
    \sqrt{\frac{2GM_p}{R_p}} > 6 \sqrt{\frac{2k_B T}{m}}
\label{eq:evaporation}
\end{equation}
From this we can see that planets are more likely to maintain atmospheres if they are larger ($R_p$) and more massive ($M_p$, which for constant density would scale as $R_p^3$) and have atmospheres that are lower in temperature ($T$) and composed of gas species with more massive particles ($m$). This dependence on the mass of the gas particles means that a planet may be able to retain heavier gas species, while lighter ones will be evaporated away. For example, in the Earth's atmosphere H$_2$ should thermally evaporate, while N$_2$ remains bound.

While the criterion above gives a first rough estimate for a planet's atmospheric stability, and its compositional dependence, in practice the evaporation of an atmosphere may be more directly shaped by the stellar wind particles and high-energy photons that impinge on a planet's atmosphere. Ultra-violet (UV) stellar photons are absorbed in the upper atmosphere, depositing heating and potentially driving a photoevaporative flow. The topics of stellar high-energy emission and atmospheric evaporation are complex and we refer the reader to other chapters in this volume (particularly those by Barman and Kubyshkina) for more detailed explanations. Nevertheless, it remains true that larger planets and those farther from their host star will be more able to retain a bound atmosphere.

Now that we have a bound gas, we can further examine how the force of gravity shapes its vertical structure. Namely, we can require that the atmosphere is in vertical hydrostatic equilibrium, with the downward force from gravity balanced by a gradient in gas pressure ($P$), to arrive at the classic expression:
\begin{equation}
    dP/dz = - \rho g
\end{equation}
For most atmospheres the gravitational acceleration ($g$) is relatively constant with altitude ($z$) and we can use the ideal gas law to relate the local gas density to its pressure and temperature: $P=\rho R T$. Here the specific gas constant ($R$) contains information about the mean molecular weight of the atmosphere ($\mu_m$), which is the average mass of a gas particle relative to the mass of a hydrogen atom ($m_H$): $R=k_B/\mu_m m_H$. We can then solve for the gas pressure in the atmosphere as a function of altitude:
\begin{equation}
    P= P_s e^{-z/H}.
\end{equation}
The pressure in the atmosphere drops off exponentially with altitude from its surface value ($P_s$) and the factor that sets how quickly it decreases is called the (pressure) scale height: 
\begin{equation}
    H=\frac{k_BT}{\mu_m m_H g}. \label{eqn:scaleheight}
\end{equation}
The physical extent of the atmosphere will be larger for gas that is hotter, lower mean molecular weight, and bound to a planet with lower gravity.

\subsection{Global energy balance and temperature estimates}

Given a planet of some mass and radius, we can see that the distance of the planet from its star will strongly shape its atmospheric properties: both through determining whether it can hold onto an atmosphere at all and also in setting the overall temperature of the atmosphere.

An estimate of a planet's atmospheric temperature can be made by assuming that the planet is in global radiative equilibrium: all of the starlight absorbed by the planet is re-radiated as thermal emission. We can then characterize that thermal emission with a temperature that we call the equilibrium temperature, $T_{\mathrm{eq}}$. Setting the starlight absorbed by the planet (on just the dayside) equal to the thermal emission from the planet (over the whole globe), we find that the planet's radius ($R_p$) cancels out and we are left with:
\begin{eqnarray}
    (\sigma_{SB} T_s^4) (R_s/a)^2 (1-A_B) \pi R_p^2 = (\sigma_{SB} T_{\mathrm{eq}}^4) 4 \pi R_p^2 \nonumber\\
    T_{\mathrm{eq}}^4 = \frac{1}{4} (1-A_B) (R_s/a)^2 T_s^4.
\end{eqnarray}
Here we have allowed for the atmosphere to reflect back some fraction of the incoming starlight, instead of absorbing it, defined by the Bond albedo ($A_B$), which is a bolometric value integrated over all wavelengths. It is convenient that the value of $T_{\mathrm{eq}}$ is set just by stellar properties (temperature and radius: $T_s$, $R_s$) and the orbital distance between the star and planet ($a$), since these are the relatively easiest parameters for us to know about exoplanet systems.

How representative the equilibrium temperature is of a planet's actual global thermal state will depend on whether there are any additional sources of heating, typically from within the planet's interior. Planets are born hot and cool over time, meaning that they are not in fact in perfect global radiative equilibrium but have some net emitted flux. There can be additional heating sources within a planet's interior, such as from gravitational differentiation of material, radiogenic heating from a solid core, or tidal heating. We can characterize any flux from the interior with another representative temperature: $T_{\mathrm{int}}=(F_{\mathrm{int}}/\sigma_{SB})^{1/4}$. Note that $T_{\mathrm{int}}$ is used just to express a flux and is often not a temperature that is actually realized within the interior. The flux from the interior may or may not be important relative to the incident stellar flux. Some useful reference points among known planets are: 
\begin{itemize}
    \item Jupiter: this planet has an internal heat flux that is roughly equal to the solar flux it receives \citep{Li2018}.
    \item Widely separated Jupiters: these directly imaged gas giants are all young (and hot) and very far from their stars, meaning that the starlight they receive is negligible compared to the flux they emit.
    \item Hot Jupiters: these planets sit several stellar radii away from their host stars, receiving $\sim10^4$ times more starlight than Jupiter. While many hot Jupiters have inflated radii, indicating much hotter interiors and higher internal heat fluxes than Jupiter, their global energy balance is still dominated by the starlight they absorb.
    \item Earth: our home planet absorbs slightly more sunlight than its thermal emission to space, the result of rapidly increasing greenhouse gases and a thermal state that has not yet had time to equilibrate \citep{Pierrehumbert}.
\end{itemize}
Since a planet's thermal state may not be well described by global balance with absorbed starlight, we also employ the quantity effective temperature, $T_{\mathrm{eff}}$, to characterize the actual thermal emission of the planet. This is set by its bolometric flux output, $F=\sigma_{SB} T_{\mathrm{eff}}^4$, and also related to the combination of equilibrium and interior fluxes: $T_{\mathrm{eff}}^4=T_{\mathrm{eq}}^4+T_{\mathrm{int}}^4$. So, as related to the example planets given above, Jupiter has $T_{\mathrm{eff}} \approx 2^{1/4} T_{\mathrm{eq}}$, widely separated Jupiters have $T_{\mathrm{eff}} \approx T_{\mathrm{int}} >> T_{\mathrm{eq}}$, and both hot Jupiters and Earth have $T_{\mathrm{eff}} \approx T_{\mathrm{eq}}$.

\subsection{Vertical temperature profiles: radiative and convective}

Up to this point we have been discussing the atmosphere as if it were described by a single temperature. In reality we expect that its vertical thermal structure will be shaped by those processes acting to transport heat through the atmosphere: radiation and convection. Within some region of the atmosphere, whichever process is more efficient will be the one that operates, and this depends on the local conditions. We can describe how well radiation passes through some material by its optical depth: a high optical depth means the material is more opaque, while a low optical depth means it is more transparent. (Later in this section we will show how optical depth appears as a quantity in radiative transfer equations.) Typically convection is more likely to occur deeper in the atmosphere, where the gas is more optically thick and where a solid surface (if present) could introduce a sharp temperature gradient, while radiative transport will take over once opacities decrease enough for the atmosphere to become translucent to radiation.

\begin{figure}
    \centering
    \includegraphics[width=\textwidth]{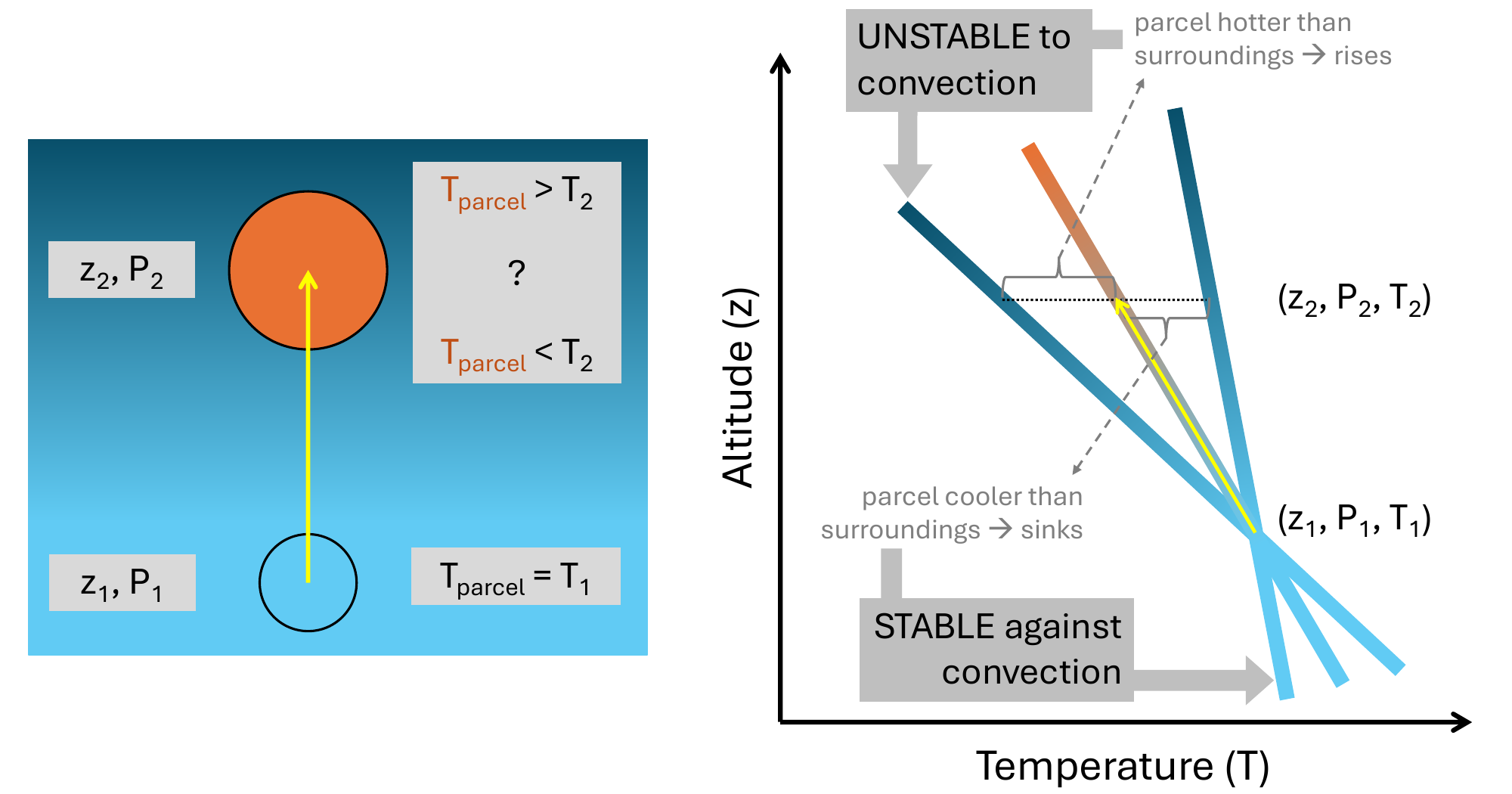}
    \caption{We can evaluate whether an atmosphere should be convective by considering the perturbation of a gas parcel to a slightly higher location in the atmosphere. We compare the temperature set by its adiabatic expansion to the temperature of its new surroundings. This determines whether it will return to its starting location (a stable atmosphere) or will continue to rise unimpeded (an unstable atmosphere). The hypothetical gas parcel is used to solve for the adiabatic temperature gradient and the criterion for convection is a comparison between this and the slope of the atmospheric temperature profile.}
    \label{fig:convection}
\end{figure}

Convection transports heat through the bulk motion of hotter gas rising into cooler regions and the settling of cooler gas down into warmer regions. For this bulk motion to occur, the atmosphere must be unstable to the perturbation of a gas parcel up (or down) from one region of the atmosphere to another; otherwise, the gas parcel will just return to its original location and this process will not vertically transport heat. We can assess the conditions for convective instability through the simple cartoon picture of a gas parcel that starts in equilibrium with its surroundings, at a location ($z_1$) with some temperature and pressure ($P_1,T_1$), and is raised up to a new location ($z_2$) at some other temperature ($T_2$). Since there is no net expansion or contraction of the atmosphere, we can apply the equation for hydrostatic equilibrium from above, so the pressure at the second location must be less than at the first ($z_2>z_1 \rightarrow P_2<P_1$). Figure \ref{fig:convection} diagrams this set-up. 

We assume the gas parcel's ascent is a quick, adiabatic process, so there is no heat exchange, but it will expand to match the new pressure conditions of its surroundings. For an ideal gas undergoing adiabatic expansion, we know from thermodynamics that the quantity $P^{1-\gamma}T^\gamma$ must remain constant, where $\gamma=c_P/c_V$ is the ratio of the specific heat capacities of the gas at constant pressure and constant volume. Remembering also from thermodynamics that $1-1/\gamma=R/c_p$, we can combine these equations to solve for how the temperature of the gas parcel will change as it is perturbed to a new height in the atmosphere, and label this as $\Gamma_d$:
\begin{equation}
    \frac{dT}{dz}=- \frac{g}{c_P} = \Gamma_d. \label{eqn:lapserate}
\end{equation}
We can see that the gas parcel will adiabatically cool when moved higher into the atmosphere. Now, if the gas parcel's new temperature is colder than its surroundings ($T_1 - \Delta z \frac{g}{c_P} < T_2$), then it is more dense and will sink back to its original location. However, if the gas parcel is warmer than its new surroundings, it is more buoyant and will continue to rise; this is an unstable perturbation that results in convection.

Now that we have derived the adiabatic temperature gradient ($\Gamma_d$, or the dry adiabatic lapse rate), we can see that what sets whether an atmosphere is unstable to convection is how its temperature gradient compares to $\Gamma_d$. If the atmosphere has a temperature gradient less than the adiabatic lapse rate ($dT/dz|_{\mathrm{atm}}<\Gamma_d$), then the atmospheric profile is ``sub-adiabatic" and stable against convection. However, if the atmospheric temperature gradient is more steep ($dT/dz|_{\mathrm{atm}}>\Gamma_d$), then it is ``super-adiabatic" and convection will be triggered. When some region of the atmosphere is convective, that process will operate efficiently to transport heat and bring the temperature profile back toward the adiabatic lapse rate. It is then a good approximation to use $\Gamma_d$ for the temperature gradient within convective regions, normalized by appropriate boundary conditions. 

If the atmosphere contains significant condensable material, then the latent heat that is released by condensation can alter the energetics of convection. Instead of the ``dry" case outlined above, if we imagine our gas parcel to be ``moist", meaning that it contains some condensable in vapor form, then as the parcel rises and cools we may expect that the material will become saturated and condense. The latent heat released will then increase the temperature of the parcel and it will become more buoyant. In other words, we should expect that convection can more easily occur in a moist atmosphere and, in fact, the moist adiabatic lapse rate is lower than the dry lapse rate. The latent heat from water clouds is an important component in the Earth's atmospheric energetics and water condensation can also play a role in shaping gas giant atmospheres, but when the mean molecular weight of the atmosphere is dominated by hydrogen and helium, there is another effect that must be taken into account. When the vapor form of a condensate is heavier than the background atmospheric gas, then a moist gas parcel will be significantly heavier than a dry one. The vertical abundance gradient of a vapor, due to its condensation, will then provide a stabilizing balance against convection and can result in super-adiabatic temperature profiles. For example, Jupiter's equatorial zone is observed to be super-adiabatic, due to a gradient in water vapor \citep{Li2024}.

Regions of the atmosphere that are stable against convection are also those where radiative transfer controls the heat transport. In this case the temperature profile is set by solving for local radiative equilibrium throughout the atmosphere, meaning that there is no net heating or cooling of the gas at any layer ($dF/dz=0$, where $F$ is the total radiative flux). This requirement constrains the thermal structure through its role in shaping the source function ($S_\nu$) in the radiative transfer equation: $dI_\nu/d\tau_\nu=S_\nu-I_\nu$, which describes how intensity ($I_\nu$) changes as it passes through some optical depth of the atmosphere ($d\tau_\nu=-\kappa_\nu ds$, where $\kappa_\nu$ is the absorption coefficient and $s$ is the physical path length). Most of these quantities are labeled with the subscript $\nu$, to explicitly indicate their dependence on the frequency of the radiation. In atmospheres we can often assume local thermodynamic equilibrium, so that the source function is given by the Planck function and set directly by the local temperature: $S_\nu=B_\nu(T)$, although this neglects scattering. The vertical thermal profile is then intricately coupled with the sources of atmospheric opacity (which set $\kappa_\nu$), since these are often temperature-dependent themselves. The absorption cross-sections of chemical species can be strong functions of temperature and pressure, but also the abundances of those species will be set by the thermal profile, as we will discuss below.

One classic solution to the intrinsically complex radiative transfer equation is that of the ``gray" atmosphere, where the absorption coefficient is assumed to be uniform, with no temperature-, pressure-, or wavelength-dependence \citep[][and many textbooks]{Eddington1916,Chandrasekhar,Mihalas}. In this case, the thermal profile can be solved for analytically, invoking other assumptions we have already introduced (such as local radiative and thermodynamic equilibria), and written using optical depth as the vertical coordinate: $T^4(\tau)=\frac{3}{4}T_{\mathrm{eff}}^4(\tau+\frac{2}{3})$. The classic gray model has long been used to approximate the atmospheric structure of stars, where the net flux ($T_{\mathrm{eff}}$) is outward through the atmosphere. However, if we assume that all planets orbit stars, then a fundamental property of planetary atmospheres is that they are irradiated. This means that instead of just an upward flux coming from the interior, there also must be an accounting for the absorption of the incident starlight. 

\begin{figure}[h]
    \centering
    \includegraphics[width=0.8\textwidth]{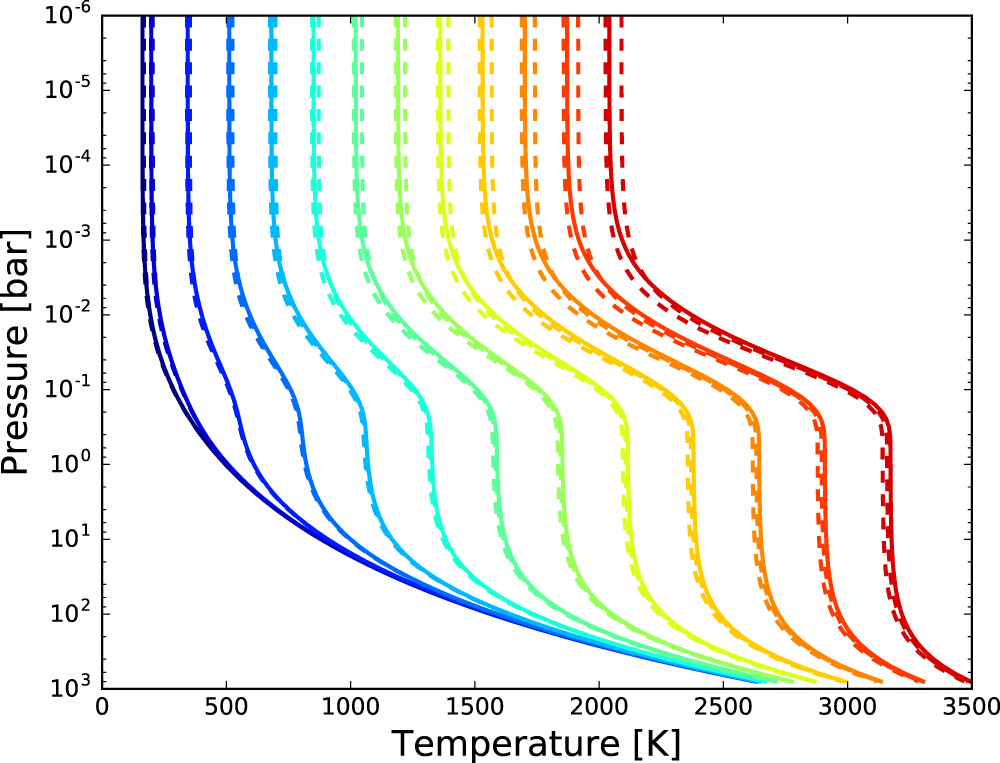}
    \caption{From \cite{Piskorz2018}, a comparison between temperature-pressure profiles from double-gray models (dashed lines) and more physically realistic models \citep[solid lines; self-consistent thermochemical-radiative-convective equilibrium models from the ScCHIMERA code,][]{Line2013}. These profiles are for a range of incident stellar fluxes, from non-irradiated (the leftmost set of profiles) to the flux at $\sim$0.03 AU from a Sun-like star. The double-gray profiles can reproduce the general shape of the more realistic profiles near the pressures where the starlight is absorbed and thermally re-emitted, but become unrealistically isothermal at low pressures. Copyrighted by the AAS and reproduced with permission.}
    \label{fig:doublegray2}
\end{figure}

The classic gray atmosphere can be extended to a slightly more complex ``double-gray" atmosphere by allowing for two uniform absorption coefficients: one that captures the absorption of starlight and one that treats the emission and absorption of thermal radiation \citep{Hubeny2003,Hansen2008,Guillot2010}. These solutions result in hotter upper atmospheres than the non-irradiated equivalent solution. For planets with temperatures much cooler than their stellar hosts, the starlight absorption coefficient can be associated with visible wavelengths, while the one that treats thermal radiation can be associated with infrared wavelengths. While the double-gray atmospheric profiles can roughly reproduce the layers where starlight is absorbed and thermal emission emerges, their main failings are that they become too isothermal in the upper atmosphere and cannot transition to convective profiles at depth. Figure \ref{fig:doublegray2} shows a comparison between double-gray profiles and those from more physically realistic models.

In general, for an atmosphere with a net transport of flux out to space, we expect that the temperature profile should decrease with increasing altitude. We see this in the gray solution for a radiative atmosphere and this trend is also mandated by the near-adiabatic profile within convective regions. However, both stars and planets can have regions of their atmospheres where the temperature instead \textit{increases} with altitude and we call these temperature inversions. Regions with increased temperatures are also those that will be emitting more thermal flux and so, for an atmospheric profile in local radiative equilibrium, they must be due to increased absorption in those layers. The Earth's stratosphere is a well known example of a region of temperature inversion, whose cause is the absorption of UV sunlight by ozone at high altitude. In this sense, temperature inversions are just a continuation of the trend where absorbed starlight results in hotter atmospheric profiles than would otherwise occur, but depends on the relative atmospheric opacity to starlight versus thermal radiation. This is exhibited in the double-gray atmospheric solution, where setting the ``visible" absorption coefficient greater than the ``infrared" one pushes the starlight to be absorbed at lower pressures and causes temperature inversions. One important consequence of temperature inversions is that they result in a planet's emission spectrum showing spectral lines in emission, instead of absorption.

The full vertical profile of the atmosphere comes from solving for radiative-convective equilibrium throughout the column. Global energy balance enforces the flux conditions at the boundaries: at the bottom boundary there is an upward flux from the interior, at the top boundary there is a downward stellar flux, and the upward flux at the top must be the sum of the two (integrated over all wavelengths). A self-consistent solution will provide, as a function of altitude: temperature, pressure, density, the abundances of chemical species, a determination of which regions are radiative and which are convective, and the emitted spectrum from the top of the atmosphere. 

\subsection{Atmospheric circulation and three-dimensional structure}

Throughout the discussion thus far we have been treating the atmosphere as if it is one-dimensional, fully described by only a vertical coordinate. This makes for a simpler picture of the planet and often this globally averaged state may be a fairly good representation of the planet overall. However, the stellar irradiation pattern on a planet is not spatially uniform and this necessarily drives atmospheric circulation. Here we provide a brief overview of the physical underpinnings of atmospheric circulation, but more detailed discussions of atmospheric dynamics can be found in other chapters within this volume, with the one by Read et al.\ focused on terrestrial worlds (including expectations for exoplanets) and the one by S\'anchez-Lavega \& Heimpel focused on Solar System gas giants. Atmospheric circulation of gas giants outside the Solar System is comprehensively reviewed in \cite{Showman2020}.

At any instant, the point on a planet that is directly facing the star (the location of ``high noon" or ``substellar" point) receives the most irradiation, with a pattern that falls off toward zero at the terminator (the dividing line between day and night). If each location on the planet had a vertical profile in local radiative-convective equilibrium, then at any given altitude there would be horizontal gradients in temperature and pressure. Since the gas in the atmosphere acts as a fluid, it should respond to these gradients by developing flow (i.e., winds) that work to minimize these gradients. However, the advective heat transport from these winds will bring each vertical profile out of local radiative equilibrium ($dF/dz \neq 0$), resulting in radiative heating or cooling that works to re-establish equilibrium, meaning that it tries to increase the horizontal temperature and pressure gradients. In other words, the atmosphere exists in a coupled physical system where differential heating drives atmospheric circulation and the circulation establishes conditions for differential heating.

From a global perspective, the stellar irradiation pattern that drives the circulation could follow the instantaneous day-night picture described above, but only if radiative heating and cooling timescales in the atmosphere are shorter than the timescale over which the stellar irradiation changes \citep[the ``diurnal" timescale, e.g., the time between successive sunrises;][]{Showman2015}. This is trivially satisfied for planets that orbit close enough to their host stars to have been tidally locked into synchronous rotation states, with permanent day- and night-sides, but is far from true for most planets in our Solar System. For planets that rotate at a faster rate than the atmosphere heats and cools (on a global scale), the stellar irradiation is effectively smeared out into a pattern set by its average over a diurnal cycle. For most of the planets in the Solar System, this results in the equator being heated more than the poles. For planets with obliquities (i.e., rotation axis tilts) greater than 54$^\circ$, such as Uranus, the poles receive more irradiation than the equator, over an orbital cycle. 

The global atmospheric circulation pattern will be set by two main components: 1) the geometry of the stellar heating and 2) the rotation rate of the planet. The first determines the direction of the overall heat transport (e.g., day to night or equator to pole), while the second prevents the winds from simply flowing in that direction. Since wind patterns are usually described within the rotating frame of the planet, this means that the Coriolis force can be seen as strongly influencing the large-scale motion of the flow. Alternatively, in an inertial reference frame the rotation rate can be seen as setting the overall angular momentum of the atmospheric flow. The faster the rotation rate of the planet, the more strongly the Coriolis force will shape its circulation pattern. For example, instead of having winds flow from the equator to the pole, in the direction of the large scale pressure gradient, the Coriolis force acts in a direction perpendicular to the winds and bends them to the east, by definition the same direction as the planet's rotation. On the Earth this breaks up direct equator-to-pole flow into three distinct regions of overturning circulation in each hemisphere, with the nearest to the equator called the Hadley Cells. Jupiter, which rotates more quickly, has instead many bands in each hemisphere, alternating between regions of eastward or westward flow.

\section{Atmospheric Chemistry}

While physics guides our understanding of the functioning of atmospheres, chemistry sets our expectations for their compositions. In practice, these two scientific analyses are not separable. As we will see, the physical conditions of the atmosphere set its chemistry, but the material in the atmosphere controls the radiative transfer (and other processes). The atmospheric physics and chemistry are linked.

\subsection{Chemical equilibrium and species abundances}

\begin{figure}
    \centering
    \includegraphics[width=0.85\textwidth]{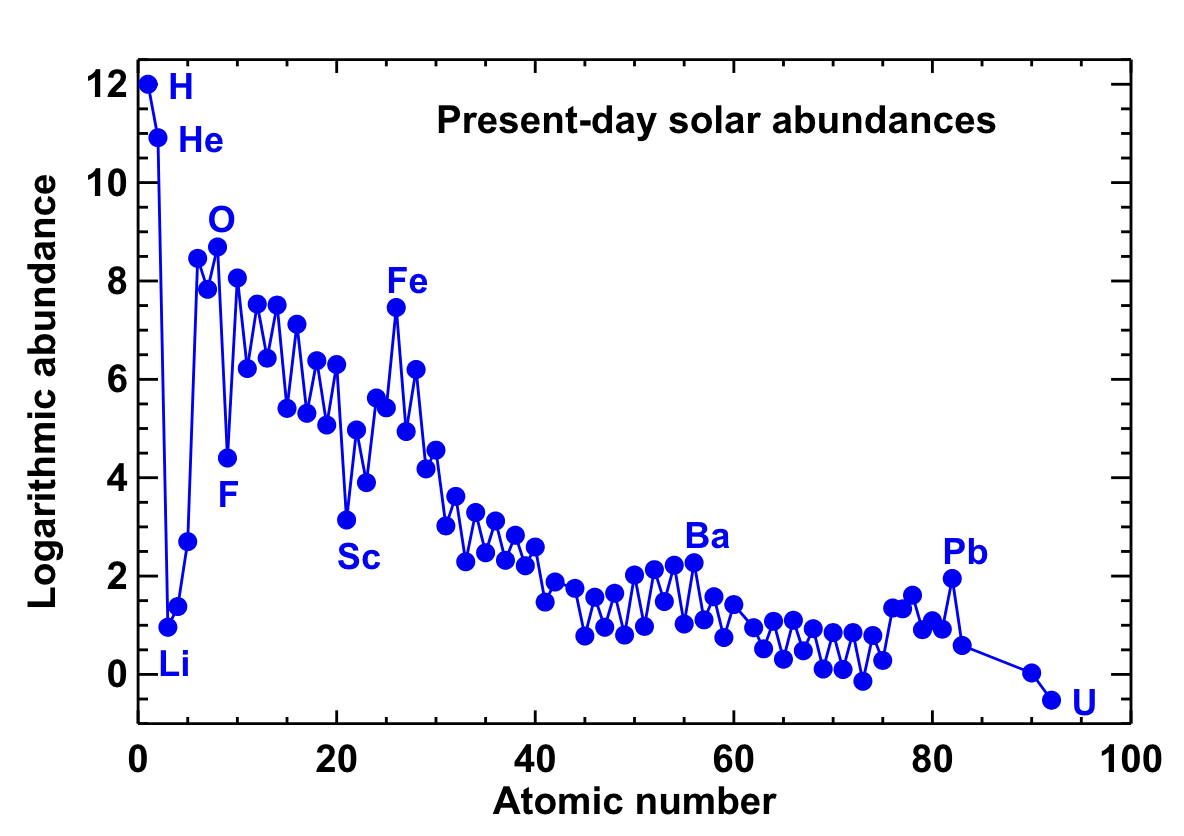}
    \caption{From \cite{Asplund2021}, the relative abundances of elements (as number per unit volume) from hydrogen through uranium in the Sun's atmosphere.}
    \label{fig:solar_abundances}
\end{figure}

Our first step in determining the composition of an atmosphere is to  know the total abundances of each element within some parcel of gas. With some expectation that these abundances are linked to the elemental abundances in a planet's host star (having formed from the same material), we often use the best measured stellar abundances as a reference: the solar elemental abundances (Figure \ref{fig:solar_abundances}). Depending on the formation pathway of a planet, we may expect its host star's elemental abundances to be more or less directly reflected in its atmosphere. Linking composition with formation is an active topic of study and reviewed in the chapter by Pudritz et al.\ within this volume. 

For a planet whose atmosphere is dominated by hydrogen and helium, i.e., a gaseous planet, we often describe its composition by its ``metallicity". This is a factor by which elements heavier than helium (the astronomer's ``metals") are enriched above their solar abundances. More specifically, in an atmospheric context the metallicity is measured as the abundance of a particular element, X, relative to hydrogen and normalized relative to the solar abundance ($\odot$), on a logarithmic scale:
\begin{equation}
    \left[\frac{\mathrm{X}}{\mathrm{H}}\right] =\log_{10}\left(\frac{(N_\mathrm{X}/N_\mathrm{H})}{(N_\mathrm{X}/N_\mathrm{H})_\odot}\right)
\end{equation}
where $N_\mathrm{X}$ is the number of atoms of the element in some unit volume of the atmosphere. If a planet has the same relative abundance of an element as the Sun, then it has a metallicity of [X/H]$=0$. When we assume that all metals are enriched by the same amount, we may use the more general notation [M/H]. Sometimes the metallicity is reported instead as a multiplicative factor so, for example, instead of saying that a planet has a metallicity of [M/H]$=0.48$, we might say that it has a metallicity 3$\times$ solar.

If we have no more detailed information, we may assume that all elements heavier than helium are enriched by the same factor, but sometimes the picture is more complicated. Depending on where a planet formed in the protoplanetary disk and the processes by which it continued to accrete material, it may be enhanced or depleted in particular elements, and this can be characterized by the abundances of two elements relative to each other. The elemental abundance ratio that often receives the most attention is the carbon-to-oxygen ratio (C/O), as these are the two most common elements after H and He (as can be seen in Figure \ref{fig:solar_abundances}), and the main carriers of carbon and oxygen are often dominant sources of opacity in atmospheres (e.g., CO, CO$_2$, CH$_4$, H$_2$O). There are also many planets, such as terrestrial worlds, where there is little to no hydrogen or helium in the atmosphere and in such cases ``metallicity" becomes a less useful quantity. For those planets, we must instead just define the abundances of different molecular species in their atmospheres.

\begin{figure}
    \centering
    \includegraphics[width=\textwidth]{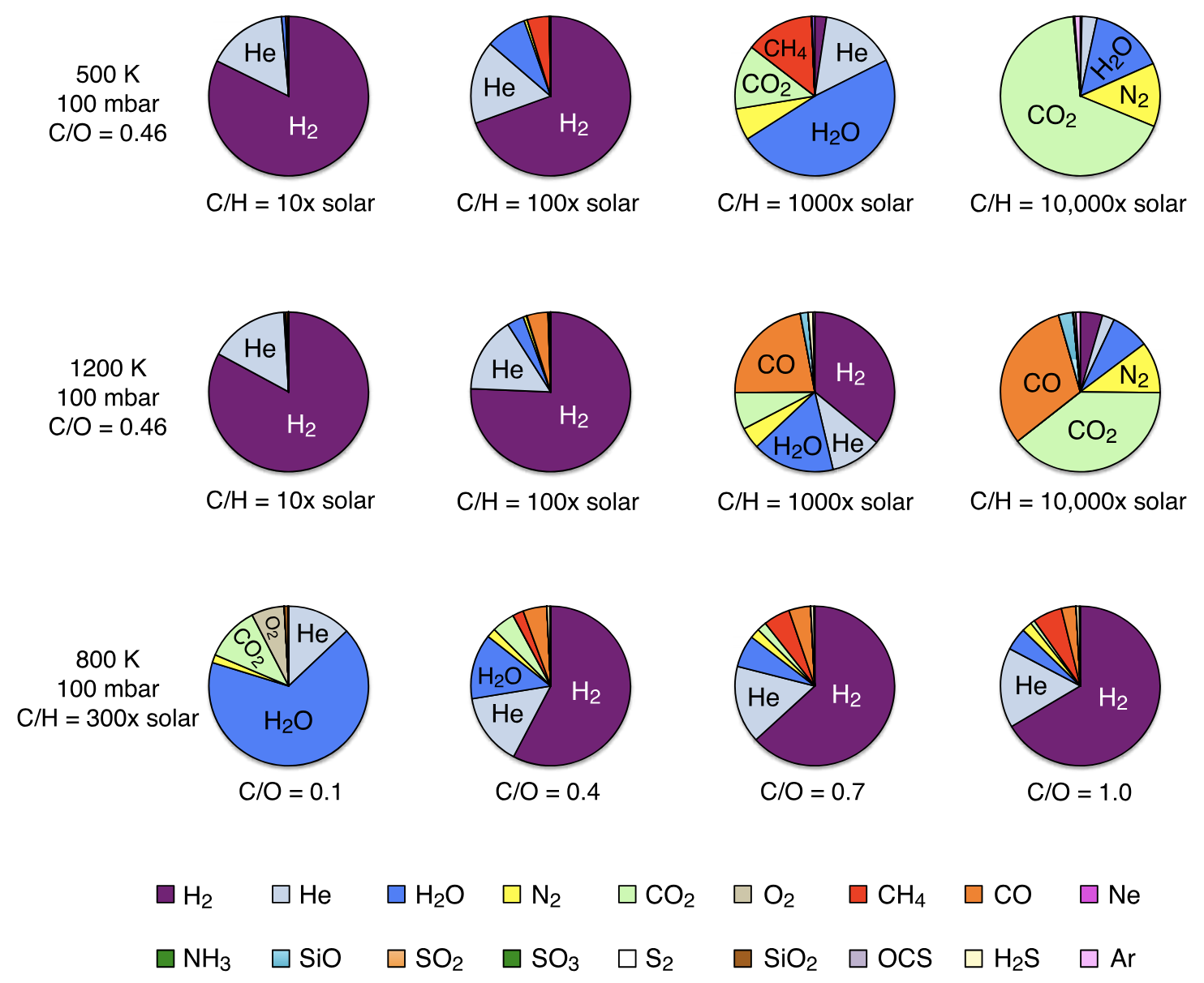}
    \caption{From \cite{Moses2013}, the thermochemical equilibrium abundances for gases with different metallicities, carbon-to-oxygen ratios, and temperatures. Copyrighted by the AAS and reproduced with permission.}
    \label{fig:piecharts}
\end{figure}

Based on the underlying, elemental composition of a gas, we can evaluate how that material should be distributed into various atomic and molecular species through the assumption of thermochemical equilibrium. This is the condition under which all of the chemical reactions converting one species to another are in balance with the equivalent back-reactions, such that the macroscopic abundances of the gas are in a steady state. Figure \ref{fig:piecharts} shows examples of thermochemical equilibrium abundances for gases with various metallicities and C/O ratios, and at different temperatures (all for $P=$100 mbar).

Thermodynamics tells us that spontaneous chemical reactions will only occur if they increase entropy, so the equilibrium state of some chemical network will only be achieved if it has enough time for reactions to maximize the entropy of the system. We can also use this condition to solve for the thermochemical equilibrium abundances in a gas: by maximizing the entropy or, equivalently, minimizing the Gibbs free energy of the system ($G=H-ST$, where $H$ is enthalpy, $S$ is entropy, and $T$ is temperature). More practically, it can be shown that minimizing $G$ is equivalent to requiring: $0 = \Delta G = \sum \mu_i dN_i $, where $N_i$ is the molar amount of species $i$ within our gas parcel and $\mu_i$ is its chemical potential, or the change to the Gibbs free energy of the system when some amount of that species is added to the system with no other parameters changed. While a somewhat circular definition, it has the benefit of incorporating all of the detailed chemistry into $\mu_i$, which is dependent on the properties of the species $i$, but also on the (fixed) temperature and pressure conditions. In practice there are tabulations of chemical potential values for a large number of species under a wide range of conditions. The boundary condition that makes it possible to solve for the abundances of all species within the chemical network of interest is that the total elemental abundances of the system are fixed. For example, the atoms of carbon may be distributed among multiple possible species, but the chemical potentials must provide an accounting for each atom. 

There are multiple codes available that use Gibbs free energy minimization to calculate thermochemical equilibrium abundances \citep[e.g.,][]{gordon94,Blecic2016,Woitke2018,Stock2022}. Once we have a predicted set of abundances for some gas in an atmosphere, we can use the radiative properties of the chemical species to determine an associated set of opacities. This then links back to the discussion of the temperature structure of the planet above: the conditions within the atmosphere are not just a result of an energetic balance at each location (from heat transport due to radiation, convection, or advection), but also the interplay between that physics and the local chemistry.

\subsection{Disequilibrium chemistry: processes and impacts}

Just as we discussed above that we may expect winds to bring temperature profiles out of radiative-convective equilibrium, we may similarly expect various processes to bring the local conditions in an atmosphere out of chemical equilibrium. We evaluate whether disequilibrium chemistry exists through a comparison of timescales: is the timescale over which our gas parcel reaches chemical equilibrium shorter or longer than the timescale over which some other process may change its physical conditions? On the Earth, the presence of life induces chemical disequilibrium in the atmosphere. More generally, one of the main processes that can cause chemical disequilibrium is the advective motion of a gas parcel: winds or turbulent motion that moves the parcel from one location in the atmosphere to another, and so from surroundings with one temperature and pressure to another. 

Here we will paint a cartoon picture of how disequilibrium chemistry may be induced within an atmosphere to present a general example of the concept, but it is important to recognize that in reality the details of atmospheric mixing and chemical reactions are more complex. We start by building upon our expectations for vertical atmospheric profiles, developed above, to note that in general the deeper regions of the atmosphere have both higher pressures and temperatures. These conditions lead to more interactions between molecules, such that we should expect overall faster chemical reaction rates. So, we might expect that the chemical timescales are shorter deeper in the atmosphere.

The vertical mixing in a planetary atmosphere is often treated as a diffusive process, such that the timescale for moving gas from one region to another is given by: $\tau_{\mathrm{mix}}=L^2/K_{zz}$, where $K_{zz}$ is called the eddy diffusion coefficient and $L$ is the length scale over which the mixing occurs, typically the atmospheric scale height or some fraction thereof. In practice, the strength of vertical mixing will depend on whether that region of the atmosphere is convective or radiative and the nature of the atmospheric circulation. For our cartoon picture, we will assume that the vertical mixing timescale is constant, or at least falls off more slowly with height than the chemical timescale.

\begin{figure}[h!]
    \centering
    \includegraphics[width=0.9\textwidth]{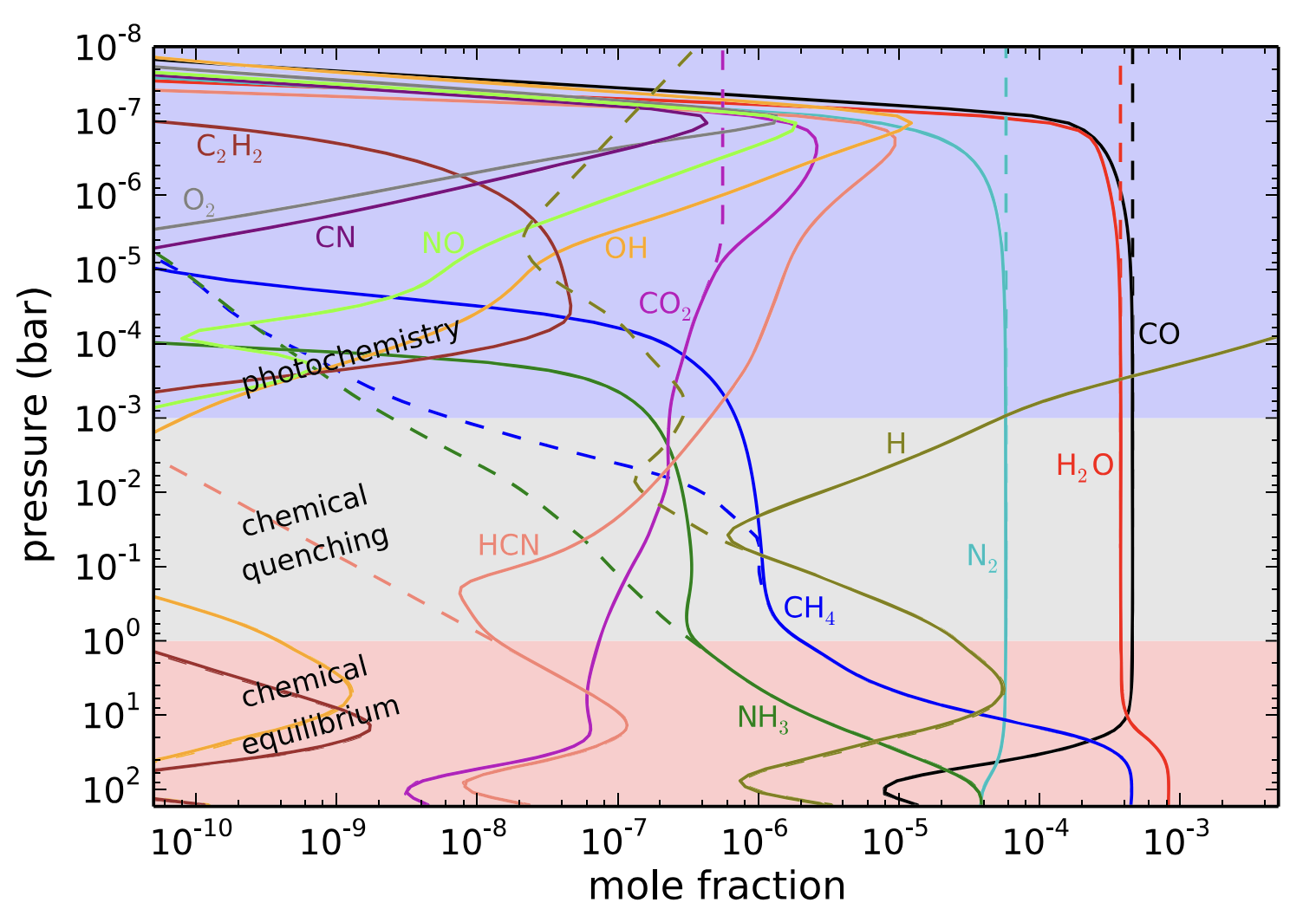}
    \caption{Originally from \cite{Agundez2014}, as visually modified in \cite{Madhu2016}, the vertical abundance profiles for various chemical species in an atmosphere. For each species the dashed line is the profile we would expect from thermochemical equilibrium, while the solid line is the profile calculated from chemical kinetics. Deep in the atmosphere (at high pressures) the abundances of species follow their equilibrium profiles, while above that disequilibrium abundances are first shaped by vertical mixing (quenching species to vertically uniform abundances) and then by photochemistry. The photodissociation of some species decreases their abundances (CO, H$_2$O, N$_2$), while also resulting in the appearance of other species (e.g., HCN, OH, O$_3$).}
    \label{fig:spaghetti}
\end{figure}

Within this picture, the deepest regions of the atmosphere will have $\tau_{\mathrm{chem}} < \tau_{\mathrm{mix}}$ and so the chemical reactions will operate fast enough for the local abundances of atomic and molecular species to be set by equilibrium conditions. Then, as we move up in the atmosphere, we will eventually cross a point above which the mixing timescale becomes faster than the chemical timescale. At this point gas parcels are mixed up into this region faster than the chemical reactions can work to bring the abundances into equilibrium with the new conditions. In this situation, where all of the gas in the upper atmosphere is very efficiently mixed, the chemical abundances will be set by chemical kinetics, which take into account the chemical reaction rates. Since the reaction rates will be the fastest at the bottom boundary of this mixed region, where $\tau_{\mathrm{chem}} = \tau_{\mathrm{mix}}$, the abundances of chemical species are ``quenched" (or fixed) to their equilibrium value at this point and then uniform throughout the mixed region above. Figure \ref{fig:spaghetti} shows an example of quenched chemical abundances, calculated from a more realistic model than our cartoon example. Since not all species have the same chemical conversion timescales, they can have quench points at different levels of the atmosphere. From observing that the abundances of atmospheric species are in disequilibrium, it may be possible to constrain the underlying atmospheric structure that sets the amount of mixing \citep{Miles2020}. The profiles in Figure \ref{fig:spaghetti} are also from a paper that studies the role of horizontal mixing on disequilibrium chemistry, as may be important in hot Jupiter atmospheres due to their strong horizontal winds and large temperature gradients \citep[e.g.,][]{Cooper2006,Lee2023,Zamyatina2023}. 

In the uppermost regions of the atmosphere another disequilibrium process becomes important: photochemistry. In this process, UV flux from the star photodissociates molecules, lowering their abundances and producing highly reactive ions and other byproduct species that would otherwise not exist. This enables new chemical reactions within the overall atmospheric network. In the Earth's atmosphere, the photodissociation of molecular oxygen (O$_2$) in the stratosphere results in the production of ozone (O$_3$). With the advent of JWST, we are now able to detect photochemically produced species in exoplanet atmospheres \citep{Tsai2023}. Figure \ref{fig:spaghetti} shows an example of how photochemistry in a planet's upper atmosphere facilitates the appearance of a wide diversity of species, by reducing the abundances of others.

\subsection{Atmospheric aerosols}

All of the discussion above has been about the composition of the gas in a planet's atmosphere; however, it is also possible for particulates of solid (or liquid) to form, which we generally refer to as aerosols. The topic of aerosols in exoplanet atmospheres is a complex, active area of research and more thoroughly reviewed in \cite{Gao2021}; here we only give a brief overview. To organize our discussion of these aerosols, we use a naming convention that has become standard within the exoplanet field and differentiates them by how they form:
\begin{description}[aerosols]
\item[Clouds]{Particulates that form when gas condenses from vapor to a solid or liquid are called clouds. Within the Earth's atmosphere, temperatures exist above and below the condensation point for water, resulting in prevalent water clouds around the planet.}
\item[Hazes]{Particulates that are produced as a result of photochemical reactions are called hazes. Smog on Earth or tholins on Titan are some of the best known examples of hazes.}
\end{description}
One commonality between clouds and hazes is that when they are present, they will change the radiative transfer through the atmosphere, influencing both the atmospheric properties and the observed planetary spectra. It is also possible (perhaps common) for both clouds and hazes to exist within an atmosphere. In this case they will influence each other both directly and indirectly. An example of a direct interaction is through the microphysics of their formation, such as the formation of ice (cloud) particles on dust (haze) grains. They indirectly influence each other through their feedback on the atmospheric chemical and physical properties.

Although cloud particles can exist in liquid or solid form, to simplify our discussion here we will use ``condensation" and ``evaporation" to describe the phase transitions out of and into gas form, respectively. At the beginning of this chapter, we discussed the conditions for the thermal evaporation of an atmosphere as when the thermal motions of a gas exceed the gravitational escape velocity. The evaporation that we are discussing now is similar in concept, in that it occurs when the thermal motions within a liquid (or solid) can overcome the intermolecular forces that maintain that phase state. While gravity provided the back-reaction against evaporation before, now the molecules in some gas may be forced to condense if they are in close enough proximity, which is determined by the ``partial pressure" of that species (i.e., the gas pressure due to that species alone). The maximum amount of a species that can exist in gas phase is quantified by its saturation vapor pressure; if the partial pressure of a gas species exceeds its saturation vapor pressure, then the excess material will be forced to condense. Different species will have different saturation vapor pressures, dependent on the details of their intermolecular bonds. However, for all species their saturation vapor pressures will increase with increasing temperature, as the thermal motion of the material can better resist condensation. A possibly familiar example of this is the appearance of dew during a cool night, formed when water vapor exceeds its saturation limit, which then evaporates during the day as the temperature rises and the saturation vapor pressure of water increases. If a condensable gas exists within a background atmosphere at some known abundance, then at every temperature-pressure location we can compare its partial pressure to its saturation vapor pressure to determine whether we expect it to condense or remain in gas form.

The condensation curve of a species is the pressure-temperature relation that delineates its phase transition between gas and either solid or liquid states. We expect clouds to form in an atmosphere if its vertical temperature profile crosses one or more condensation curves. In the deeper, hotter region of an atmosphere we may expect some species to exist in vapor form, but if gas containing that species is moved to a higher altitude, across that species' condensation curve, it should form cloud particles. The formation of clouds is intrinsically linked with the atmospheric circulation: ``moist" gas must be moved from warmer to cooler locations. When condensates are sequestered in clouds, they are also necessarily depleted in their gas phase abundance above the cloud layer, which is important to take into account when converting from observed species abundances to underlying elemental abundances \citep[e.g.,][]{Visscher2010,Helling2016}. While for many planets it is primarily vertical motion that forms clouds, for planets with large temperature gradients around the planet it can also be horizontal motion that brings material across condensation curves; this can even be the dominant mode of cloud formation \citep[e.g.,][]{Powell2024}.

Once formed, the cloud particles will then want to gravitationally settle, falling deeper into the atmosphere, where they might then evaporate back into gaseous form. This sets up a competition between vertical motion mixing cloud particles up and gravitational settling pulling them down; the balance point between these will determine how vertically extended the clouds are (i.e., how physically thick the cloud layer is). However, the microphysics of cloud formation quickly makes the picture more complicated: the details of how the cloud particles form and grow, over what timescales and with what efficiency, and the resulting size distribution of particles (which then sets their gravitational settling timescales). Vertical cloud distributions therefore require complex microphysical modeling or must be simplified through parameterization \citep[e.g.,][]{Ackerman2001}.

Since hazes are formed from chemical reactions driven by high-energy stellar irradiation, we may expect them to primarily exist on the dayside of a planet and in its upper atmosphere, where the photodissociation occurs. However, the microphysics of hazes is similarly complicated, set by the details of chemical formation pathways and particle growth timescales, underlying circulation patterns that can advect the hazes through changing atmospheric conditions, and radiative feedback with the atmospheric structure \citep[e.g.,][]{Steinrueck2023}. Haze particles will be subject to the same gravitational settling as cloud particles, but expectations for the destruction of hazes is far less clear than the evaporation of clouds \citep[e.g.,][]{Yu2021_hazesurfaces}. It remains an active area of investigation what the compositions of hazes even should be for different atmospheric conditions \citep[e.g.,][]{He2020,Moran2020}.

\section{Comparing Terrestrial and Gas Giant Atmospheres}

Within our Solar System, the planets fall neatly into two types: gas giants and terrestrial worlds. A gas giant has an atmosphere composed primarily of hydrogen and helium, which we assume it has retained from its initial formation, as a primordial atmosphere. In contrast, we can use the balance between thermal and escape velocities discussed above to show that terrestrial planets are unlikely to have gravitationally retained any significant hydrogen or helium from the protoplanetary disk in which they formed. Instead, terrestrial planets have secondary atmospheres, composed of material that is released from within the planet's interior through outgassing during its lifetime. These different paths of formation and evolution result in two main differences between their atmospheres:
\begin{enumerate}
    \item While the composition of a gas giant atmosphere is dominated by molecular hydrogen and helium, terrestrial planet atmospheres are instead made of higher mean molecular weight material. The volatile species outgassed from the planet's interior will be primarily carriers of carbon, nitrogen, and oxygen (such as CO$_2$, N$_2$, and H$_2$O).
    \item The atmosphere of a gas giant is just the outermost regions of the gaseous envelope that dominates its mass, while the atmosphere of a terrestrial planet is a small fraction of its total mass and sits above a ``solid" surface. The surface of the Earth is in some places solid and in some liquid, but there is nevertheless a phase transition at the surface, which is importantly different from the more continuous transition between the atmosphere and interior of gas giants.
\end{enumerate}

While a clean distinction between terrestrial and gas giants exists within our Solar Sytem, the most prevalent type of planet discovered to date are the intermediate-sized planets known as super-Earths and sub-Neptunes. The formation pathways, global structures, and atmospheric compositions of these common planets remain topics of active study \citep[see the review by][]{Bean2021}. From just the masses and radii of these planets alone, many are consistent with a variety of possible global compositions, resulting in large uncertainties in their atmospheric compositions and whether there exists a surface under a thin atmosphere or one buried deep under a thick envelope. For the sake of this chapter, we note that these complex intermediate-case planets exist and then ignore them in the discussion below, focusing instead on the simpler delineation between classic terrestrial planets and gas giants.

\subsection{Atmospheric compositions dominated, or not, by H$_2$ and He}

For atmospheres dominated by molecular hydrogen and helium, these species will set the overall properties of the gas, such as the mean molecular weight and specific gas constant, while the other elements are primarily important in that they compose the main sources of opacity. Jupiter and Saturn are our best examples of the H/He-dominated regime. If we compare the atmospheres of these two planets, we might initially expect the atmosphere of Saturn to have a higher mean molecular weight, since it seems to have a higher overall metallicity (10$\times$ solar, compared to 3$\times$ solar). However, Saturn's atmosphere is depleted in helium and this influences the mean molecular weight more strongly than the metallicity enrichment, resulting in Saturn actually having a lower atmospheric mean molecular weight than Jupiter. Both Uranus and Neptune have even more highly metal-enriched atmospheres, at $\sim100\times$ solar, which means that C, N, and O have become only a factor of 10 less abundant than H and He. This does result in a higher mean molecular weight for these atmospheres, but they are still composed of less than a few percent of species other than H$_2$ and He (primarily CH$_4$).

Where is the transition between H/He-dominated atmospheres and ones dominated by heavier molecules? While the examples of Uranus and Neptune show that [M/H] = 2 still places an atmosphere within the H/He-dominated regime, once metallicity is increased to 1000$\times$ solar, the chemical equilibrium abundances for most atmospheric conditions predict that heavier molecules become dominant (see Figure \ref{fig:piecharts}). This demonstrates that at high enough values, ``metallicity" becomes a poor descriptor of the compositions of these atmospheres and we instead may describe them based on which molecules are the main constituents, for example a CO$_2$-dominated atmosphere like Venus or a hypothetical water-world with a steam-dominated atmosphere.

\subsection{The presence or lack of a surface}

Whether or not a planet has a surface, there are three types of interactions with the interior that happen at the bottom boundary of its atmosphere: 1) compositional mixing, 2) angular momentum exchange, and 3) heat transport.

A planet's global composition may be more or less evenly distributed between its atmosphere and interior. For terrestrial planets, the origin of their atmospheres in the first place is the outgassing of volatile species from their interiors, but they can also continue to have material exchange with the interior through surface interactions. Condensates can rain out of the atmosphere onto the surface or evaporate from the surface into the atmosphere. Dust grains can get lofted up into the atmosphere or settle back to the surface, a prominent changing feature of Mars' atmosphere. The Earth's carbon cycle enables CO$_2$ in the atmosphere to be sequestered below the surface, into the ocean and crustal material, and then released back into the atmosphere. The wider diversity of interactions available among the terrestrial exoplanet population is reviewed by \cite{Lichtenberg2024}. The compositional mixing between the atmosphere and deeper interior of giant planets may involve the differentiation of material, such as the formation and rain out of helium droplets (which can also deplete neon), but the details are uncertain, even in the Solar System gas giants \citep[see the review by][]{Guillot2022}.

When a surface exists under an atmosphere, it acts as a strong boundary condition on the atmospheric winds: they must go to zero. The surface therefore imposes a strong drag on the circulation, which is simple in concept but more complex in the details of the turbulence induced in the planetary boundary layer. In contrast, a gas giant's atmosphere does not have an obvious frictional surface as its bottom boundary, but instead there can be fluid interactions deep into the planet. Eventually, at high enough pressures the electrical properties of molecular hydrogen change and it becomes conductive (i.e., can carry a current). At this point magnetohydrodynamic effects should drag the fluid into solid body rotation, effectively damping the wind speeds \citep{Liu2008}. The depth of the jets in Jupiter's atmosphere supports this picture \citep{Kaspi2018}. For hot Jupiter planets, where the radiative atmosphere extends much deeper than on Jupiter, the angular momentum interaction between the atmosphere and interior is potentially more complex \citep{Carone2020}.

We introduced atmospheres as the boundary layer between the planet's interior and the rest of the universe. The atmosphere mediates the absorption of starlight and the planet's thermal emission to space, which is a combination of re-processed starlight and flux from the interior. The flux from the interior will be transmitted through the atmosphere regardless of whether the planet has a surface or not, but the re-processing of the starlight can work differently if there is a surface, depending on whether or not the atmosphere above it is optically thick to starlight. In the case that the atmosphere is optically thick and the starlight never reaches the surface (e.g., Venus), then the starlight is absorbed and re-processed as thermal emission within the atmosphere itself, the same as if the planet were a gas giant.

For a terrestrial planet with an atmosphere that is optically thin to the starlight (such as the Earth), then the surface will absorb or reflect whatever starlight reaches it. This heats the surface, which will then re-radiate in the infrared. Many molecules are strong sources of infrared opacity, so an atmosphere that is optically thin at visible wavelengths can often be optically thick in the infrared. We can treat this type of atmosphere as a gray atmosphere, described above, where the re-processed starlight just becomes a flux up from the bottom boundary and the absorption coefficient applies to the infrared radiation. At some point the density in the atmosphere drops low enough that it becomes optically thin and the temperature at that point will set the thermal emission to space ($T=T_{\mathrm{eff}}$). For a planet in global radiative equilibrium, and with negligible interior flux, the emitted flux must equal the absorbed flux from the starlight (and so $T_{\mathrm{eff}}=T_{\mathrm{eq}}$). Importantly, since the solution for a gray atmosphere has temperature decrease with altitude, this necessarily means that the surface of the planet is hotter than the equilibrium temperature. More generally, for more physically realistic atmospheric profiles, the same situation still holds true and is called the greenhouse effect: a terrestrial planet with an atmosphere that is optically thin to incoming starlight but optically thick to the re-emitted thermal radiation will have a surface that is hotter than it would be in the absence of the atmosphere.

\section{Conclusions}

We have now come the end of our relatively brief tour through fundamental concepts in atmospheric physics and chemistry, but there is a much larger and even more interesting landscape to explore. Thanks to the diversity of conditions realized among exoplanets we get to take the science anchored by Solar System atmospheres and expand it into a wide variety of new regimes. These studies are also not purely theoretical. The last two dozen years have seen the development of new techniques for measuring more and more detailed properties of exoplanet atmospheres. New data from JWST are dramatically advancing our ability to piece apart the physical and chemical complexity of atmospheres across the galaxy. We anticipate another similar jump in understanding when the Extremely Large Telescopes come online. While atmospheric physics and chemistry provide rich topics of study on their own, they are also foundational to related avenues of inquiry. If we are going to look in exoplanet atmospheres for signatures of planet formation processes or biologically induced disequilibrium chemistry, we must first have confidence that we understand the physical and chemical conditions within which those markers exist.

\section{Cross-References}
\begin{itemize}
\item{Radiative transfer for exoplanet atmospheres}
\item{Retrievals}
\item{Planetary evaporation through evolution}
\item{Planetary Atmospheres Through Time: Effects of Mass Loss and Thermal Evolution}
\item{Connecting Planetary Composition with Formation}
\item{Composition and Chemistry of the Atmospheres of Terrestrial Planets: Venus, the Earth, Mars, and Titan}
\item{Temperature, Clouds, and Aerosols in the Terrestrial Bodies of the Solar System}
\item{Temperature, Clouds, and Aerosols in Giant and Icy Planets}
\item{Atmospheric Dynamics of Terrestrial Planets}
\item{Atmospheric Dynamics of Giants and Icy Planets}
\end{itemize}

\begin{acknowledgement}
I thank Natasha Batalha for her initial contributions to defining the scope of this chapter, Ted Bergin for double-checking my chemistry, and Jonathan Fortney for providing valuable and constructive feedback. I also thank the Simons Foundation for gifting me with the chance to spend more time thinking about atmospheric physics (and chemistry).
\end{acknowledgement}

\bibliographystyle{spbasicHBexo}  
\bibliography{HBexoTemplatebib} 

\begin{thebibliography}{44}
\providecommand{\natexlab}[1]{#1}
\providecommand{\url}[1]{{#1}}
\providecommand{\urlprefix}{URL }
\expandafter\ifx\csname urlstyle\endcsname\relax
  \providecommand{\doi}[1]{DOI~\discretionary{}{}{}#1}\else
  \providecommand{\doi}{DOI~\discretionary{}{}{}\begingroup \urlstyle{rm}\Url}\fi
\providecommand{\eprint}[2][]{\url{#2}}

\bibitem[{{Ackerman} and {Marley}(2001)}]{Ackerman2001}
{Ackerman} AS {Marley} MS (2001) {Precipitating Condensation Clouds in Substellar Atmospheres}. \apj 556(2):872--884

\bibitem[{{Ag{\'u}ndez} et~al.(2014){Ag{\'u}ndez}, {Parmentier}, {Venot}, {Hersant}, and {Selsis}}]{Agundez2014}
{Ag{\'u}ndez} M, {Parmentier} V, {Venot} O, {Hersant} F {Selsis} F (2014) {Pseudo 2D chemical model of hot-Jupiter atmospheres: application to HD 209458b and HD 189733b}. \aap 564:A73

\bibitem[{{Asplund} et~al.(2021){Asplund}, {Amarsi}, and {Grevesse}}]{Asplund2021}
{Asplund} M, {Amarsi} AM {Grevesse} N (2021) {The chemical make-up of the Sun: A 2020 vision}. \aap 653:A141

\bibitem[{{Bean} et~al.(2021){Bean}, {Raymond}, and {Owen}}]{Bean2021}
{Bean} JL, {Raymond} SN {Owen} JE (2021) {The Nature and Origins of Sub-Neptune Size Planets}. Journal of Geophysical Research (Planets) 126(1):e06639

\bibitem[{{Blecic} et~al.(2016){Blecic}, {Harrington}, and {Bowman}}]{Blecic2016}
{Blecic} J, {Harrington} J {Bowman} MO (2016) {TEA: A Code Calculating Thermochemical Equilibrium Abundances}. \apjs 225(1):4

\bibitem[{{Carone} et~al.(2020){Carone}, {Baeyens}, {Molli{\`e}re}, {Barth}, {Vazan}, {Decin}, {Sarkis}, {Venot}, and {Henning}}]{Carone2020}
{Carone} L, {Baeyens} R, {Molli{\`e}re} P et~al. (2020) {Equatorial retrograde flow in WASP-43b elicited by deep wind jets?} \mnras 496(3):3582--3614

\bibitem[{{Chandrasekhar}(1960)}]{Chandrasekhar}
{Chandrasekhar} S (1960) {Radiative transfer}

\bibitem[{{Cooper} and {Showman}(2006)}]{Cooper2006}
{Cooper} CS {Showman} AP (2006) {Dynamics and Disequilibrium Carbon Chemistry in Hot Jupiter Atmospheres, with Application to HD 209458b}. \apj 649(2):1048--1063

\bibitem[{{Eddington}(1916)}]{Eddington1916}
{Eddington} AS (1916) {On the radiative equilibrium of the stars}. \mnras 77:16--35

\bibitem[{{Gao} et~al.(2021){Gao}, {Wakeford}, {Moran}, and {Parmentier}}]{Gao2021}
{Gao} P, {Wakeford} HR, {Moran} SE {Parmentier} V (2021) {Aerosols in Exoplanet Atmospheres}. Journal of Geophysical Research (Planets) 126(4):e06655

\bibitem[{{Gordon} and {McBride}(1994)}]{gordon94}
{Gordon} S {McBride} BJ (1994) Computer program for calculation of complex chemical equilibrium compositions and applications. NASA Reference Publication 1311

\bibitem[{{Greene} et~al.(2023){Greene}, {Bell}, {Ducrot}, {Dyrek}, {Lagage}, and {Fortney}}]{Greene2023}
{Greene} TP, {Bell} TJ, {Ducrot} E et~al. (2023) {Thermal emission from the Earth-sized exoplanet TRAPPIST-1 b using JWST}. \nat 618(7963):39--42

\bibitem[{{Guillot}(2010)}]{Guillot2010}
{Guillot} T (2010) {On the radiative equilibrium of irradiated planetary atmospheres}. \aap 520:A27

\bibitem[{{Guillot} et~al.(2022){Guillot}, {Fletcher}, {Helled}, {Ikoma}, {Line}, and {Parmentier}}]{Guillot2022}
{Guillot} T, {Fletcher} LN, {Helled} R et~al. (2022) {Giant Planets from the Inside-Out}. arXiv e-prints arXiv:2205.04100

\bibitem[{{Hansen}(2008)}]{Hansen2008}
{Hansen} BMS (2008) {On the Absorption and Redistribution of Energy in Irradiated Planets}. \apjs 179(2):484--508

\bibitem[{{He} et~al.(2020){He}, {H{\"o}rst}, {Lewis}, {Yu}, {Moses}, {McGuiggan}, {Marley}, {Kempton}, {Morley}, {Valenti}, and {Vuitton}}]{He2020}
{He} C, {H{\"o}rst} SM, {Lewis} NK et~al. (2020) {Haze Formation in Warm H$_{2}$-rich Exoplanet Atmospheres}. \psj 1(2):51

\bibitem[{{Helling} et~al.(2016){Helling}, {Lee}, {Dobbs-Dixon}, {Mayne}, {Amundsen}, {Khaimova}, {Unger}, {Manners}, {Acreman}, and {Smith}}]{Helling2016}
{Helling} C, {Lee} E, {Dobbs-Dixon} I et~al. (2016) {The mineral clouds on HD 209458b and HD 189733b}. \mnras 460(1):855--883

\bibitem[{{Hubeny} et~al.(2003){Hubeny}, {Burrows}, and {Sudarsky}}]{Hubeny2003}
{Hubeny} I, {Burrows} A {Sudarsky} D (2003) {A Possible Bifurcation in Atmospheres of Strongly Irradiated Stars and Planets}. \apj 594(2):1011--1018

\bibitem[{{Kaspi} et~al.(2018){Kaspi}, {Galanti}, {Hubbard}, {Stevenson}, {Bolton}, {Iess}, {Guillot}, {Bloxham}, {Connerney}, {Cao}, {Durante}, {Folkner}, {Helled}, {Ingersoll}, {Levin}, {Lunine}, {Miguel}, {Militzer}, {Parisi}, and {Wahl}}]{Kaspi2018}
{Kaspi} Y, {Galanti} E, {Hubbard} WB et~al. (2018) {Jupiter{\textquoteright}s atmospheric jet streams extend thousands of kilometres deep}. \nat 555(7695):223--226

\bibitem[{{Kreidberg} et~al.(2019){Kreidberg}, {Koll}, {Morley}, {Hu}, {Schaefer}, {Deming}, {Stevenson}, {Dittmann}, {Vanderburg}, {Berardo}, {Guo}, {Stassun}, {Crossfield}, {Charbonneau}, {Latham}, {Loeb}, {Ricker}, {Seager}, and {Vanderspek}}]{Kreidberg2019}
{Kreidberg} L, {Koll} DDB, {Morley} C et~al. (2019) {Absence of a thick atmosphere on the terrestrial exoplanet LHS 3844b}. \nat 573(7772):87--90

\bibitem[{{Lee} et~al.(2023){Lee}, {Tsai}, {Hammond}, and {Tan}}]{Lee2023}
{Lee} EKH, {Tsai} SM, {Hammond} M {Tan} X (2023) {A mini-chemical scheme with net reactions for 3D general circulation models. II. 3D thermochemical modelling of WASP-39b and HD 189733b}. \aap 672:A110

\bibitem[{{Li} et~al.(2024){Li}, {Allison}, {Atreya}, {Brueshaber}, {Fletcher}, {Guillot}, {Li}, {Lunine}, {Miguel}, {Orton}, {Steffes}, {Waite}, {Wong}, {Levin}, and {Bolton}}]{Li2024}
{Li} C, {Allison} M, {Atreya} S et~al. (2024) {Super-adiabatic temperature gradient at Jupiter's equatorial zone and implications for the water abundance}. \icarus 414:116028

\bibitem[{{Li} et~al.(2018){Li}, {Jiang}, {West}, {Gierasch}, {Perez-Hoyos}, {Sanchez-Lavega}, {Fletcher}, {Fortney}, {Knowles}, {Porco}, {Baines}, {Fry}, {Mallama}, {Achtergerg}, {Simon}, {Nixon}, {Orton}, {Dyudina}, {Ewald}, and {Schmude}}]{Li2018}
{Li} L, {Jiang} X, {West} RA et~al. (2018) {Less absorbed solar energy and more internal heat for Jupiter}. Nature Communications 9:3709

\bibitem[{{Lichtenberg} and {Miguel}(2024)}]{Lichtenberg2024}
{Lichtenberg} T {Miguel} Y (2024) {Super-Earths and Earth-like Exoplanets}. arXiv e-prints arXiv:2405.04057

\bibitem[{{Line} et~al.(2013){Line}, {Wolf}, {Zhang}, {Knutson}, {Kammer}, {Ellison}, {Deroo}, {Crisp}, and {Yung}}]{Line2013}
{Line} MR, {Wolf} AS, {Zhang} X et~al. (2013) {A Systematic Retrieval Analysis of Secondary Eclipse Spectra. I. A Comparison of Atmospheric Retrieval Techniques}. \apj 775(2):137

\bibitem[{{Liu} et~al.(2008){Liu}, {Goldreich}, and {Stevenson}}]{Liu2008}
{Liu} J, {Goldreich} PM {Stevenson} DJ (2008) {Constraints on deep-seated zonal winds inside Jupiter and Saturn}. \icarus 196(2):653--664

\bibitem[{{Madhusudhan} et~al.(2016){Madhusudhan}, {Ag{\'u}ndez}, {Moses}, and {Hu}}]{Madhu2016}
{Madhusudhan} N, {Ag{\'u}ndez} M, {Moses} JI {Hu} Y (2016) {Exoplanetary Atmospheres{\textemdash}Chemistry, Formation Conditions, and Habitability}. \ssr 205(1-4):285--348

\bibitem[{{Mihalas}(1978)}]{Mihalas}
{Mihalas} D (1978) {Stellar atmospheres}

\bibitem[{{Miles} et~al.(2020){Miles}, {Skemer}, {Morley}, {Marley}, {Fortney}, {Allers}, {Faherty}, {Geballe}, {Visscher}, {Schneider}, {Lupu}, {Freedman}, and {Bjoraker}}]{Miles2020}
{Miles} BE, {Skemer} AJI, {Morley} CV et~al. (2020) {Observations of Disequilibrium CO Chemistry in the Coldest Brown Dwarfs}. \aj 160(2):63

\bibitem[{{Moran} et~al.(2020){Moran}, {H{\"o}rst}, {Vuitton}, {He}, {Lewis}, {Flandinet}, {Moses}, {North}, {Orthous-Daunay}, {Sebree}, {Wolters}, {Kempton}, {Marley}, {Morley}, and {Valenti}}]{Moran2020}
{Moran} SE, {H{\"o}rst} SM, {Vuitton} V et~al. (2020) {Chemistry of Temperate Super-Earth and Mini-Neptune Atmospheric Hazes from Laboratory Experiments}. \psj 1(1):17

\bibitem[{{Moses} et~al.(2013){Moses}, {Line}, {Visscher}, {Richardson}, {Nettelmann}, {Fortney}, {Barman}, {Stevenson}, and {Madhusudhan}}]{Moses2013}
{Moses} JI, {Line} MR, {Visscher} C et~al. (2013) {Compositional Diversity in the Atmospheres of Hot Neptunes, with Application to GJ 436b}. \apj 777(1):34

\bibitem[{{Pierrehumbert}(2010)}]{Pierrehumbert}
{Pierrehumbert} RT (2010) {Principles of Planetary Climate}

\bibitem[{{Piskorz} et~al.(2018){Piskorz}, {Buzard}, {Line}, {Knutson}, {Benneke}, {Crockett}, {Lockwood}, {Blake}, {Barman}, {Bender}, {Deming}, and {Johnson}}]{Piskorz2018}
{Piskorz} D, {Buzard} C, {Line} MR et~al. (2018) {Ground- and Space-based Detection of the Thermal Emission Spectrum of the Transiting Hot Jupiter KELT-2Ab}. \aj 156(3):133

\bibitem[{{Powell} and {Zhang}(2024)}]{Powell2024}
{Powell} D {Zhang} X (2024) {Two-Dimensional Models of Microphysical Clouds on Hot Jupiters I: Cloud Properties}. arXiv e-prints arXiv:2404.08759

\bibitem[{{Seager}(2010)}]{Seager}
{Seager} S (2010) {Exoplanet Atmospheres: Physical Processes}

\bibitem[{{Showman} et~al.(2015){Showman}, {Lewis}, and {Fortney}}]{Showman2015}
{Showman} AP, {Lewis} NK {Fortney} JJ (2015) {3D Atmospheric Circulation of Warm and Hot Jupiters}. \apj 801(2):95

\bibitem[{{Showman} et~al.(2020){Showman}, {Tan}, and {Parmentier}}]{Showman2020}
{Showman} AP, {Tan} X {Parmentier} V (2020) {Atmospheric Dynamics of Hot Giant Planets and Brown Dwarfs}. \ssr 216(8):139

\bibitem[{{Steinrueck} et~al.(2023){Steinrueck}, {Koskinen}, {Lavvas}, {Parmentier}, {Zieba}, {Tan}, {Zhang}, and {Kreidberg}}]{Steinrueck2023}
{Steinrueck} ME, {Koskinen} T, {Lavvas} P et~al. (2023) {Photochemical Hazes Dramatically Alter Temperature Structure and Atmospheric Circulation in 3D Simulations of Hot Jupiters}. \apj 951(2):117

\bibitem[{{Stock} et~al.(2022){Stock}, {Kitzmann}, and {Patzer}}]{Stock2022}
{Stock} JW, {Kitzmann} D {Patzer} ABC (2022) {FASTCHEM 2 : an improved computer program to determine the gas-phase chemical equilibrium composition for arbitrary element distributions}. \mnras 517(3):4070--4080

\bibitem[{{Tsai} et~al.(2023){Tsai}, {Lee}, {Powell}, {Gao}, {Zhang}, {Moses}, {H{\'e}brard}, {Venot}, {Parmentier}, {Jordan}, {Hu}, {Alam}, {Alderson}, {Batalha}, {Bean}, {Benneke}, {Bierson}, {Brady}, {Carone}, {Carter}, {Chubb}, {Inglis}, {Leconte}, {Line}, {L{\'o}pez-Morales}, {Miguel}, {Molaverdikhani}, {Rustamkulov}, {Sing}, {Stevenson}, {Wakeford}, {Yang}, {Aggarwal}, {Baeyens}, {Barat}, {de Val-Borro}, {Daylan}, {Fortney}, {France}, {Goyal}, {Grant}, {Kirk}, {Kreidberg}, {Louca}, {Moran}, {Mukherjee}, {Nasedkin}, {Ohno}, {Rackham}, {Redfield}, {Taylor}, {Tremblin}, {Visscher}, {Wallack}, {Welbanks}, {Youngblood}, {Ahrer}, {Batalha}, {Behr}, {Berta-Thompson}, {Blecic}, {Casewell}, {Crossfield}, {Crouzet}, {Cubillos}, {Decin}, {D{\'e}sert}, {Feinstein}, {Gibson}, {Harrington}, {Heng}, {Henning}, {Kempton}, {Krick}, {Lagage}, {Lendl}, {Lothringer}, {Mansfield}, {Mayne}, {Mikal-Evans}, {Palle}, {Schlawin}, {Shorttle}, {Wheatley}, and {Yurchenko}}]{Tsai2023}
{Tsai} SM, {Lee} EKH, {Powell} D et~al. (2023) {Photochemically produced SO$_{2}$ in the atmosphere of WASP-39b}. \nat 617(7961):483--487

\bibitem[{{Visscher} et~al.(2010){Visscher}, {Lodders}, and {Fegley}}]{Visscher2010}
{Visscher} C, {Lodders} K {Fegley} J Bruce (2010) {Atmospheric Chemistry in Giant Planets, Brown Dwarfs, and Low-mass Dwarf Stars. III. Iron, Magnesium, and Silicon}. \apj 716(2):1060--1075

\bibitem[{{Woitke} et~al.(2018){Woitke}, {Helling}, {Hunter}, {Millard}, {Turner}, {Worters}, {Blecic}, and {Stock}}]{Woitke2018}
{Woitke} P, {Helling} C, {Hunter} GH et~al. (2018) {Equilibrium chemistry down to 100 K. Impact of silicates and phyllosilicates on the carbon to oxygen ratio}. \aap 614:A1

\bibitem[{{Yu} et~al.(2021){Yu}, {He}, {Zhang}, {H{\"o}rst}, {Dymont}, {McGuiggan}, {Moses}, {Lewis}, {Fortney}, {Gao}, {Kempton}, {Moran}, {Morley}, {Powell}, {Valenti}, and {Vuitton}}]{Yu2021_hazesurfaces}
{Yu} X, {He} C, {Zhang} X et~al. (2021) {Haze evolution in temperate exoplanet atmospheres through surface energy measurements}. Nature Astronomy 5:822--831

\bibitem[{{Zamyatina} et~al.(2023){Zamyatina}, {H{\'e}brard}, {Drummond}, {Mayne}, {Manners}, {Christie}, {Tremblin}, {Sing}, and {Kohary}}]{Zamyatina2023}
{Zamyatina} M, {H{\'e}brard} E, {Drummond} B et~al. (2023) {Observability of signatures of transport-induced chemistry in clear atmospheres of hot gas giant exoplanets}. \mnras 519(2):3129--3153

\end{thebibliography}

\end{document}